\documentclass[pre,twocolumn]{revtex4}
\usepackage[utf8]{inputenc}
\usepackage{amssymb,amsfonts,amsmath}
\usepackage{grffile,float,xcolor,xspace}
\usepackage{natbib}
\usepackage{graphicx}
\begin{document}

\title{Advent of Extreme Events in Predator Populations}
%in  Lotka-Volterra coupling}
\author{Sudhanshu Shekhar Chaurasia$^1$, Umesh Kumar Verma$^{1,2}$ and Sudeshna Sinha$^1$}
\affiliation{$^1$ Indian Institute of Science Education and Research Mohali, Knowledge City, Sector 81, Manauli PO 140 306, India\\
$^2$ Indian Institute of Science Education and Research Tirupati, Tirupati, 517507, India}
\date{\today}
\begin{abstract}
We study the dynamics of a ring of patches with vegetation-prey-predator populations, coupled through interactions of the Lotka-Volterra type. We find that the system yields aperiodic, recurrent and rare explosive bursts of predator density in a few isolated spatial patches from time to time. Further, the collective predator biomass also exhibits sudden uncorrelated occurrences of large deviations as the coupled system evolves. The maximum value of the predator population in a patch, as well as the maximum value of the predator biomass, increases with coupling strength. These trends are further  corroborated by fits to  Generalized Extreme Value distributions, where the location and scale factor of the distribution increases markedly with coupling strength, indicating the crucial role of coupling interactions in the generation of  extreme events. These results indicate how occurrences of extremely large predator populations can emerge in coupled population dynamics, and in a more general context they suggest a generic class of deterministic nonlinear systems that can naturally exhibit extreme events.
\end{abstract}
\maketitle

%%%%%%%%%%%%%%%%%%%%%%%%%%%%%%%%%%%%%%%%%%%%%%%%%%%%%%%%%%%

%\section{Introduction}

%\bigskip
%\bigskip
%\bigskip

\noindent
{\bf Introduction}\\

Due to their huge impact in phenomena that range from  traffic jams to weather disturbances,
the existence of extreme events has triggered much research interest   \cite{extreme}. An extreme event can be considered as one where a state variable (or variables) in an engineered or natural system exhibits large deviations from the average, i.e. the system is interrupted by sudden excursions to values that are significantly different from the mean value, with such deviations being aperiodic, recurrent and rare. Typically  an extreme event can be said to have occurred if a variable is several standard deviations away from the mean, and such unusually large values signal occurrences of catastrophic significance. Examples of such extreme events are found in weather patterns \cite{weather}, ocean waves \cite{ocean}, financial crashes \cite{market}, black-outs in power grid networks \cite{powergrid1,powergrid2} and optical systems \cite{optical}.

%% The para below has to be re-written - as it is a cut-and-paste from our Chaos paper. But the content can be the same

The search for generic mechanisms that naturally yield such extreme events is an issue of vital importance for the basic understanding of complex systems, as well as real world applications \cite{promit}. Efforts to obtain extreme events typically involve stochastic models \cite{satya1,satya2}, such as a recent random walk model of transport on networks \cite{santhanam}. In the arena of deterministic dynamical systems, there have also been a few recent studies on extreme event generation in coupled systems. Such systems have typically been composed of diffusively coupled individual units that are excitable systems, which are capable of self-generating large deviations \cite{ulrike1,ulrike2,ulrike3}. It is important however to find broad coupling classes that can provide mechanisms to induce extreme events in dynamical systems that are not capable of generating such extreme events  in isolation. Unearthing such deterministic systems would offer non-trivial examples of extreme events arising from interactions \cite{promit}, rather than intrinsic or noise-driven large deviations in the states of the constituent systems.

Here we explore the emergence of extreme events in a ring of patches with vegetation-prey-predator populations, coupled through interactions of the Lotka-Volterra type. Unlike many earlier models yielding extreme events, our model has no stochastic environmental influences or sources of random fluctuations, in either the state variables or the parameters determining the dynamics of isolated sites. 
%Nor are the dynamical constituents of our network excitable units, capable of self-producing ``pulse-like'' behaviour or ``spikes''. 
Rather, we present a new scenario for the emergence of extreme events in both space and time, in a system of populations coupled through generic Lotka-Volterra type interactions, suggesting a generic coupling class that can naturally yield extreme events in interactive deterministic nonlinear systems.

%%%%%%%%%%%%%%%%%%%%%%%%%%%%%%%%%%%%%%%%%%%%%%%%%%%%%%%%%%%

%\section{Model}

%% Add references to Blasius and to some of the papers referred to in his paper

%In 1999, Blasius et. al. published a paper, where they have proposed 

\bigskip
\bigskip

\noindent
{\bf Coupled Population Model}\\

In this work we consider a model for the population fluctuation in snowshoe hare and the Canadian lynx, that fits observed data well \cite{blasius}. The dynamical equations describing the time evolution of the coupled system of vegetation (denoted by $u$), prey (denoted by $v$) and predator (denoted by $w$) populations is given by the following functions $f(u,v,w)$, $g(u,v,w)$ and $h(u,v,w)$: 
\begin{eqnarray}
\label{eqn_blasius}
\dot{u}&=&f(u,v,w) \ = \ a u -\alpha_1f_1(u,v)\nonumber \\
\dot{v}&=&g(u,v,w) \ = \ -b v+\alpha_1f_1(u,v)-\alpha_2f_2(v,w) \nonumber \\
\dot{w}&=&h(u,v,w) \ = \ -c(w-w^\star)+\alpha_2f_2(v,w)
\end{eqnarray}
where the functions $f_1(u,v)$ and $f_2(v,w)$ describe two types of coupling. The functional form of $f_1(u,v)$ is $uv/(1+ku)$, i.e. it is a Holling type II term representing the coupling of vegetation ($u$) and prey ($v$) with coupling strength $\alpha_1$. The second function $f_2(v,w) = v w$, is the Lotka-Volterra term representing coupling of prey ($v$) and predator ($w$) with coupling strength $\alpha_2$. In this work we chose the parameters to be $a=1.0$, $b=1.0$, $c=10.0$, $w^\star=0.006$, $\alpha_1=0.5$, $\alpha_2=1.0$ and $k=0.05$ \cite{blasius}.
%%The parameters are the ones used by Blasius - right?
%which are kept fixed throughout the work.\\

\begin{figure}[H]
	\centering
	\includegraphics[trim=8cm 5cm 8cm 2cm, clip, width=0.275\linewidth]{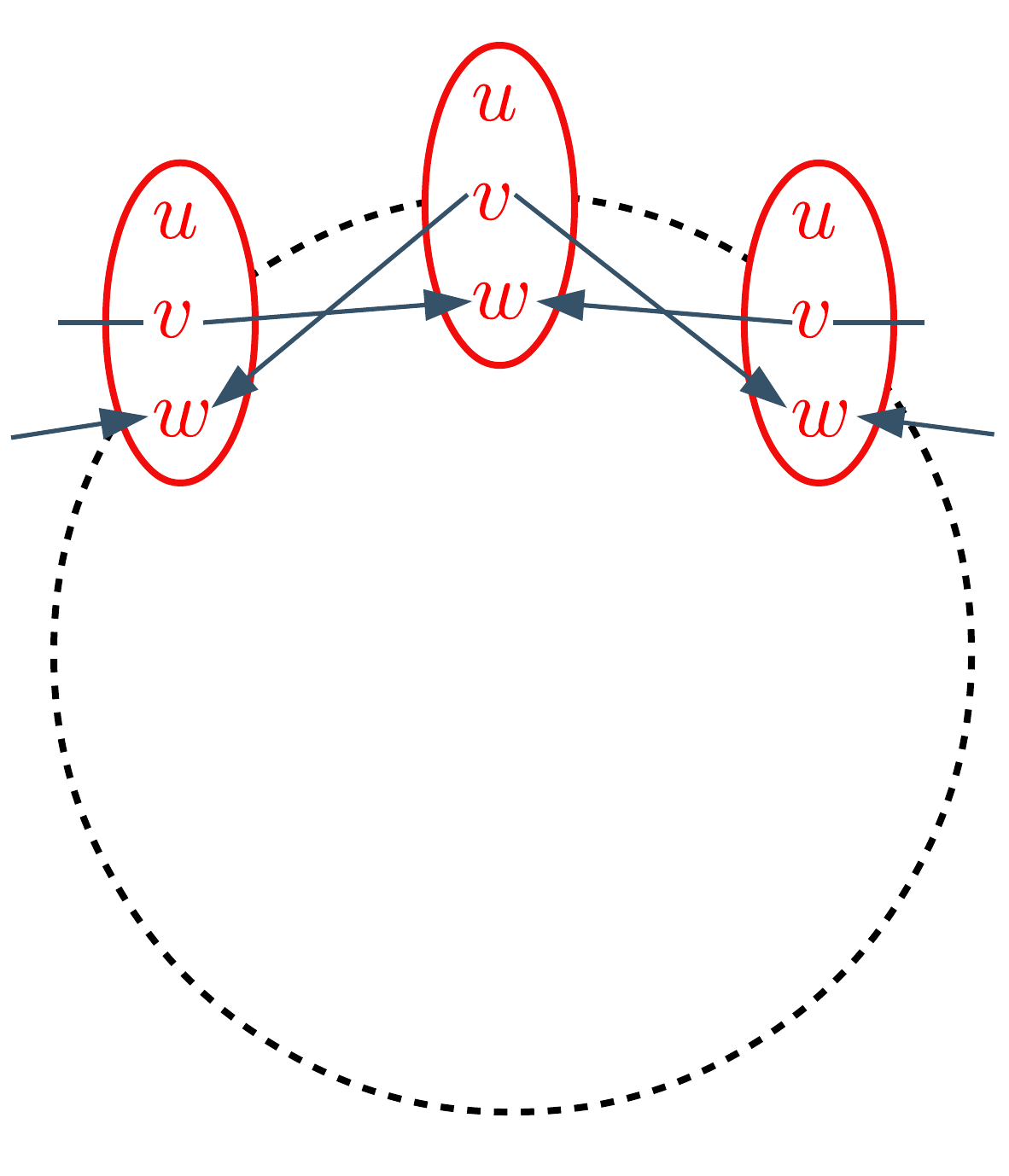}
	\caption{Schematic diagram of the interaction of a patch with nearest neighbour patches, in a ring of population patches.}
	\label{fig_schematic}
\end{figure}

Now we expand the scope of the model above to mimic a collection of such population patches, consisting of vegetation, preys and predators. The populations in a patch interact with other nearby patches in such a way that the predator of one patch can attack prey in neighbouring patches. Fig~\ref{fig_schematic} shows a schematic diagram of the interaction of a patch with nearby patches. Specifically, we consider a ring of such patches, with each patch indexed spatially by $i$, $i=1,\dots N$, where $N$ is the total number of patches in the ring. The populations of vegetation, prey and predator in patch $i$ is represented by $u_i$, $v_i$ and $w_i$  respectively. 
%with periodic boundary conditions because of ring network.

The form of the predator-prey interaction between neighbouring patches is of the Lotka-Volterra type and is given by the following set of dynamical equations: %Eqn~\ref{eqn_model}
\begin{eqnarray}
	\label{eqn_model}
	\dot{u_i}&=&f(u_i,v_i,w_i) \notag\\
	\dot{v_i}&=&g(u_i,v_i,w_i) \ - \ \frac{C}{2} \{ \ v_i w_{i-1} + v_i w_{i+1} \}\\
	\dot{w_i}&=&h(u_i,v_i,w_i) \ + \ \frac{C}{2} \ \{ w_i v_{i-1} + w_i v_{i+1} \} \notag
\end{eqnarray}
The coupling constant $C$ reflects the strength of interaction among patches, and in this work we focus on this crucial parameter. 
%The additional coupling term in prey populations ($v_i$) shows  reduction due to predator interaction, resulting in increment of predator population ($w_i$).

%%%%%%%%%%%%%%%%%%%%%%%%%%%%%%%%%%%%%%%%%%%%%%%%%%%%%%%%%%%

%\section{Spatial distribution of predator population of patches}

\bigskip
\bigskip

\noindent
{\bf Spatial distribution and temporal evolution of the population densities in the network}\\

The first quantity of relevance in this system is the {\em local population density} in the patches, and their temporal fluctuations. We will focus on the deviation of the local population densities from the mean value, i.e., we will look for the emergence of extreme events in the local patches, as evident in explosive population densities at specific spatial locations.

\begin{figure}[htb]
	\centering
    \includegraphics[width=0.49\linewidth,height=0.75\linewidth
    ]{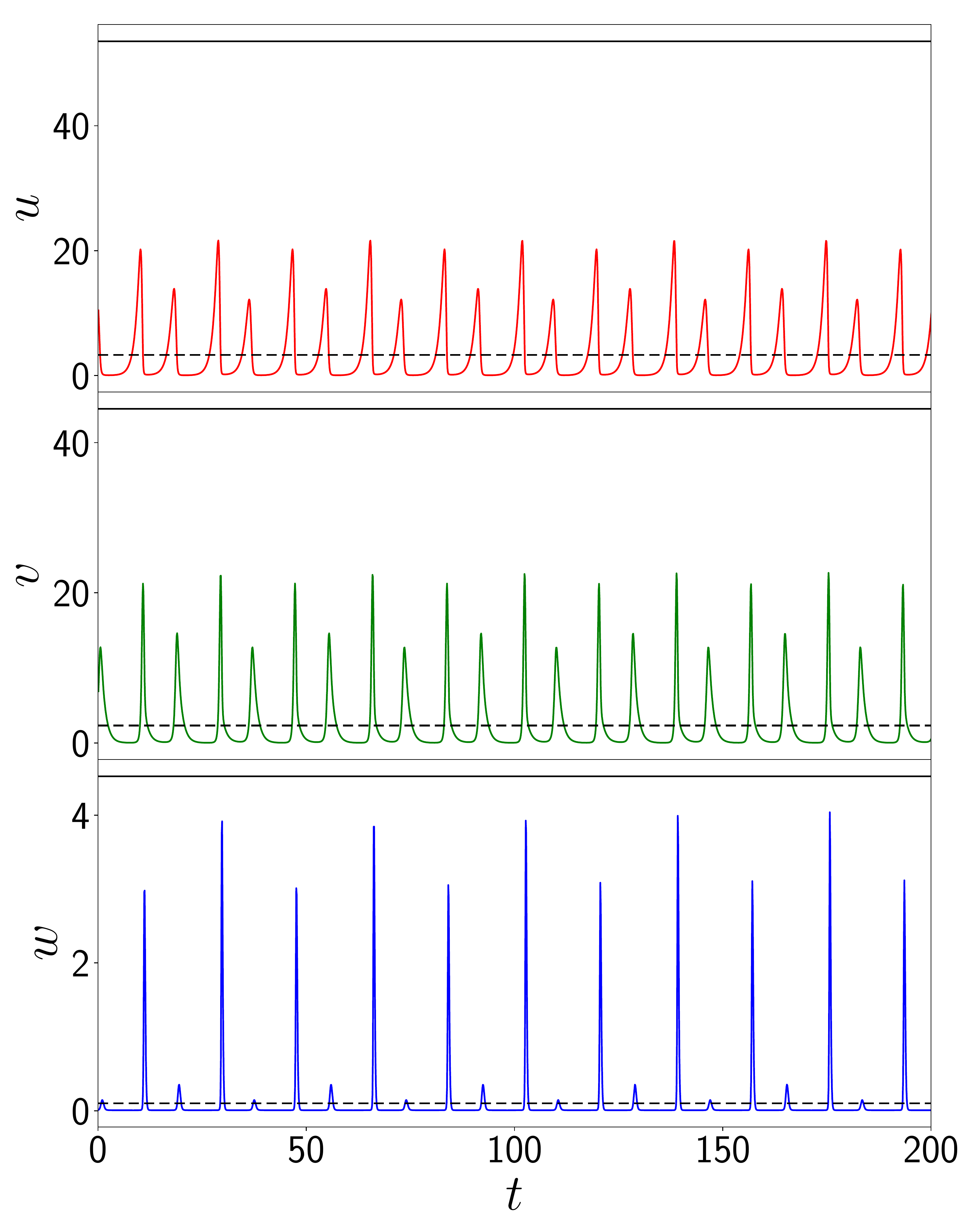}
    \includegraphics[width=0.49\linewidth,height=0.75\linewidth
    ]{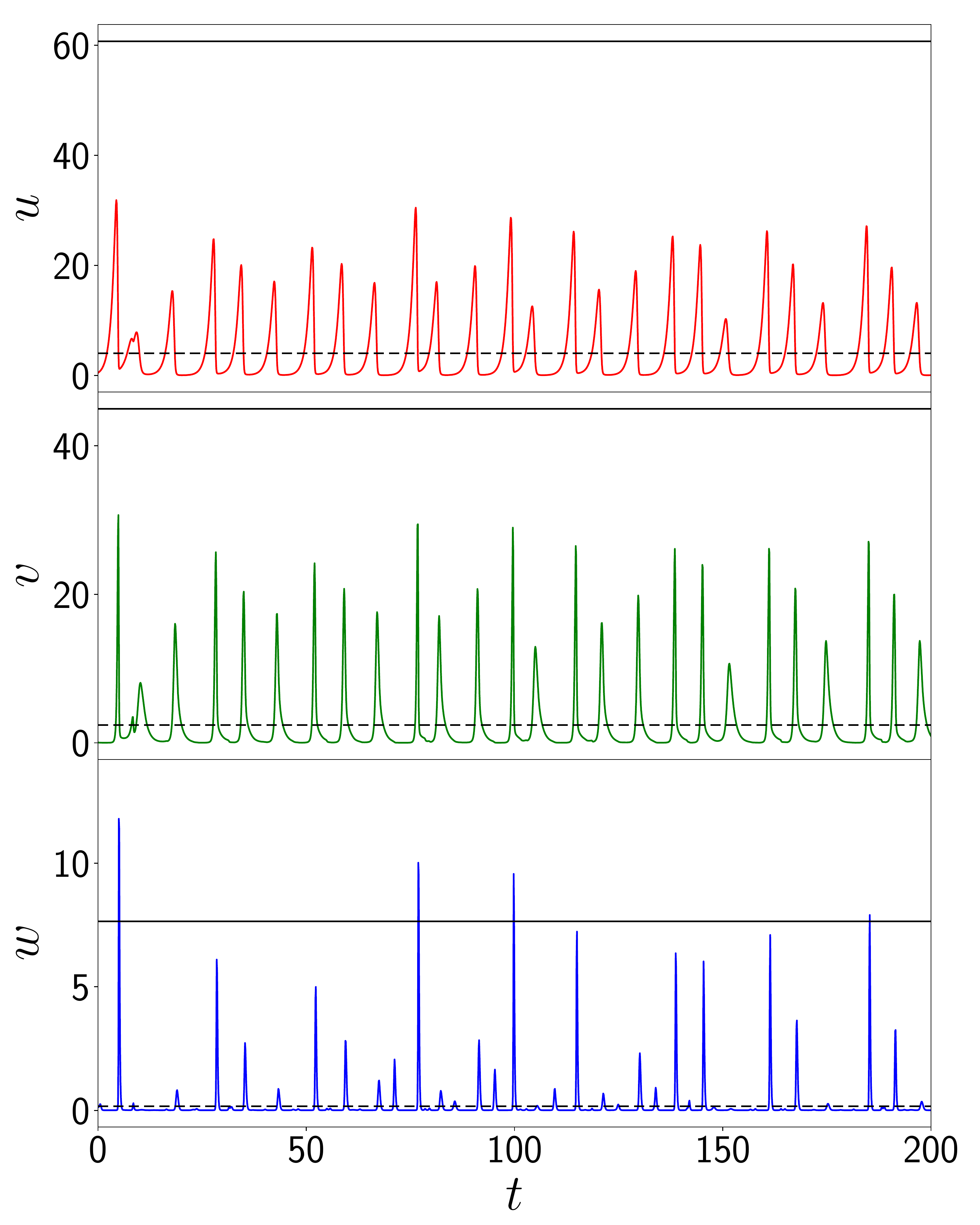}
    \caption{Time evolution of the vegetation $u$, prey $v$ and predator $w$ populations of one representative patch, for the following cases: (left) uncoupled patches and (right) patches coupled to neighbouring patches, with coupling strength $C=1.0$ in Eqn.~\ref{eqn_model}. The black dashed and black solid lines represent the mean ($\mu$) and ten times the standard deviation $\sigma$ from the mean (i.e. $\mu+10\sigma$) respectively.}
    \label{fig_population_in_space_uvw}
\end{figure}

\begin{figure}[htb]
	\centering
    \includegraphics[width=0.85\linewidth,height=0.4675\linewidth]{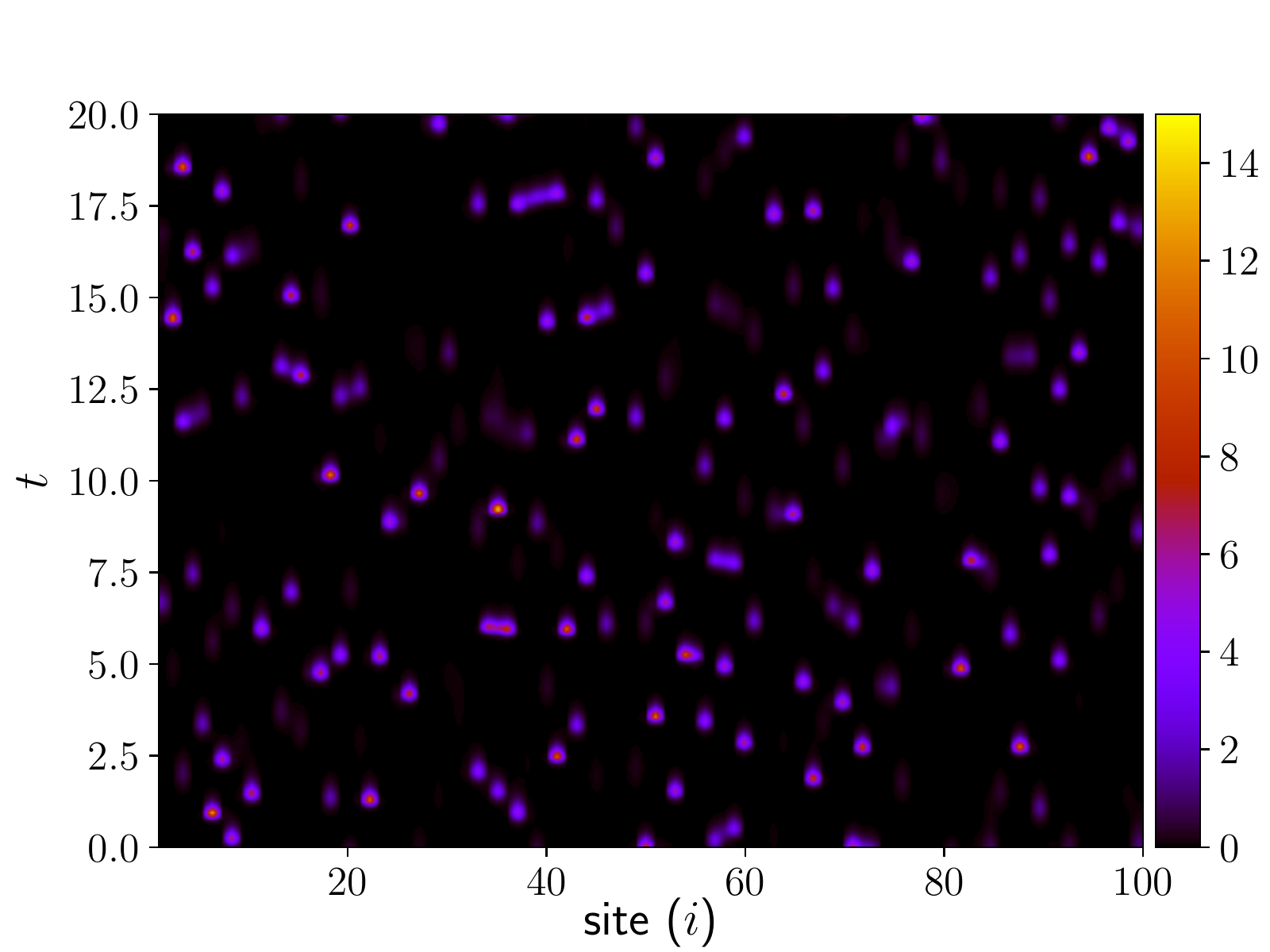}
    \caption{Space-time density plot of the predator population $w_i(t)$ at site $i=1, \dots 100$, for the case of patches coupled to neighbouring patches, with coupling strength $C=1.0$ in Eqn.~\ref{eqn_model}.}
    \label{fig_population_in_space_and_time}
\end{figure}

Fig.~\ref{fig_population_in_space_uvw} shows the time evolution of the vegetation, prey and predator populations of one representative patch from the network at high coupling strength $C=1.0$. We observe that, while the population densities are mostly confined to low values, the evolution is punctuated by sudden boosts to very high values. For instance, local predator population densities can shoot up more than $10$ standard deviations away from the mean value. This is evident in the lower right panel of Fig.~\ref{fig_population_in_space_uvw}, where one can see  instances where $w$ exceeds the $10 \sigma$ threshold. The instants at which these large fluctuations occur are relatively rare and completely {\em uncorrelated} in time and space.
%, and the maximum population densities \cite{balki1,balki2} attained in the course of the network dynamics increase monotonically with coupling strength. This is especially true of the predator population, where the maximum predator population in the coupled case can be seen to be over three times that in the uncoupled patches (cf. Fig.~\ref{fig_population_in_space_uvw} lower panels).
These large deviations are also clearly evident through the space-time plot of the evolution of predator populations in different patches in the network, shown in Fig.~\ref{fig_population_in_space_and_time}, where the extreme values of predator populations are visible as sparse bright randomly located dots in the figure.

Note that if consider a lower threshold (e.g. $\mu + 4 \sigma$), the vegetation, prey and predator all exhibit events that exceed the threshold, with prey and predator populations yielding above-threshold events even when uncoupled. However significantly, such occurrences in uncoupled patches are strictly {\em periodic}, and stem from the intrinsic pulse-like solutions of the constituent patches. 
%So while exploring extreme events in local patches, in space and in time, we consider the $4\sigma$ threshold.

\begin{figure}[htb]
	\centering
    \includegraphics[width=0.4875\linewidth,height=0.45\linewidth]{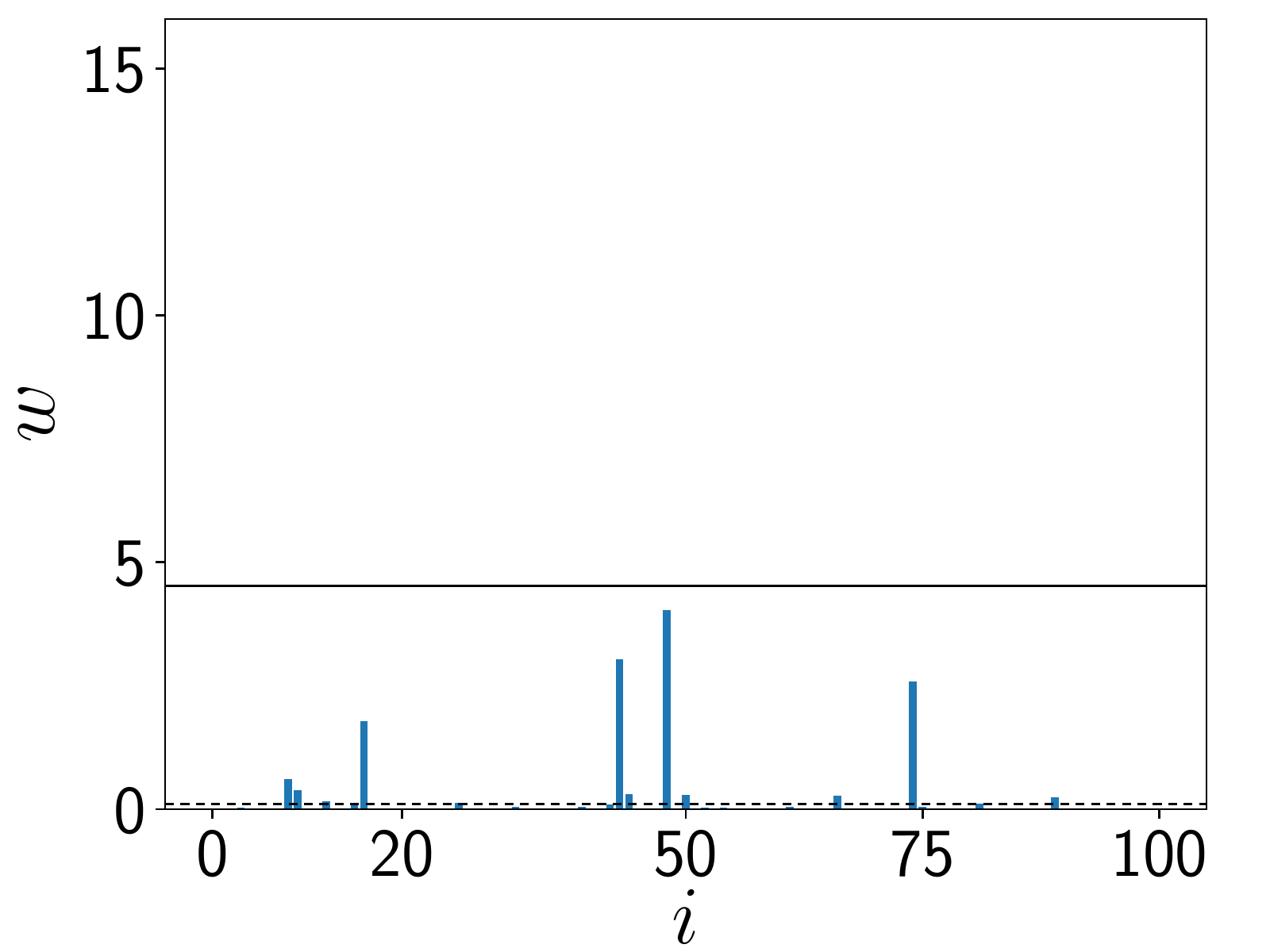}
        \includegraphics[width=0.4875\linewidth,height=0.45\linewidth]{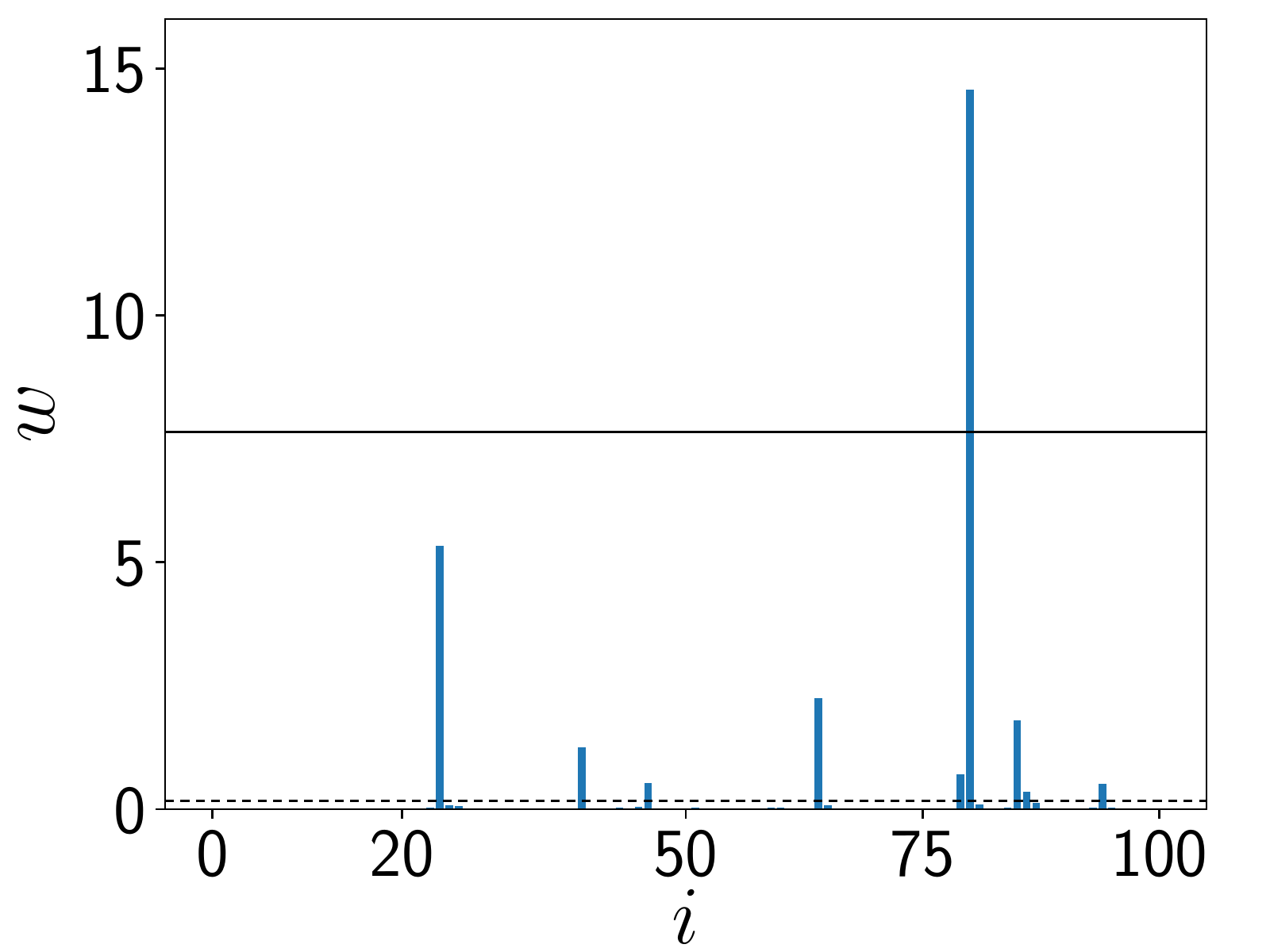}
    \caption{Spatial distribution of the predator population $w_i$  ($i=1, \dots 100$), at a representative instant of time, for the following cases: (left) uncoupled patches, and (right) patches coupled to neighbouring patches, with coupling strength $C=1.0$ in Eqn.~\ref{eqn_model}. The black dashed and black solid lines represent the mean ($\mu$) and ten times the standard deviation $\sigma$ from the mean (i.e. $\mu+10\sigma$) respectively. Note that the predator population $w_i$ in the most patches, as well as the spatial average of $w$ at an instant of time, is so low that it is barely visible in the figures.}
    \label{fig_population_in_space}
\end{figure}

In Fig.~\ref{fig_population_in_space} we display the spatial distributions of the predator populations $w_i$ for the patches $i=1, \dots N$ at a representative instant of time, for the case of uncoupled patches and coupled patches. It is clearly seen that there are a few patches in the coupled case that grow explosively and have a predator population much larger than the mean value. These results qualitatively suggest that the coupling of population patches give rise to extreme predator populations at a small number of spatial locations, analogous to an {\em extreme event in space.} Such catastrophic events are rare in space, and occurs only at a couple of sites, but signal significant damage as they may entail serious control costs.

%\begin{figure}[H]
%	\centering
%	\includegraphics[width=1\linewidth,height=0.6\linewidth]{biomass_N100_cp_0.pdf}
%	\includegraphics[width=1\linewidth,height=0.6\linewidth]{biomass_N100_cp_1.pdf}
%	\caption{Time evolution of the biomass $b_w (t)$ of the predator population, for the following cases: (top) uncoupled patches and (bottom) patches coupled to neighbouring patches, with coupling strength $C=1.0$ in Eqn.~\ref{eqn_model}. The black and red lines represent the mean ($\mu$) and four times the standard deviation $\sigma$ from the mean (i.e. $\mu+4\sigma$) respectively.}
%	\label{fig_biomass_with_time}
%\end{figure}

\bigskip
\bigskip

\noindent
\textbf{Global maximum of predator populations}\\

Now we quantitatively estimate the maximum vegetation, prey and predator population densities in a patch, denoted by $u_{max}$, $v_{max}$ and $w_{max}$ respectively, attained in the course of the network dynamics \cite{balki1,balki2}. We estimate this by finding the global maximum of $u_i$, $v_i$ and $w_i$, for $i=1, \dots N$ (where $N$ is sufficiently large), sampled over a time interval $T$ (where $T$ is much longer than the intrinsic oscillation period). This will help us gauge the magnitude of the extreme event and its relation to coupling strengths. Here we present results with $T=50$, with no loss of generality.

\begin{figure}[htb]
	\centering
	\includegraphics[width=0.6\linewidth]{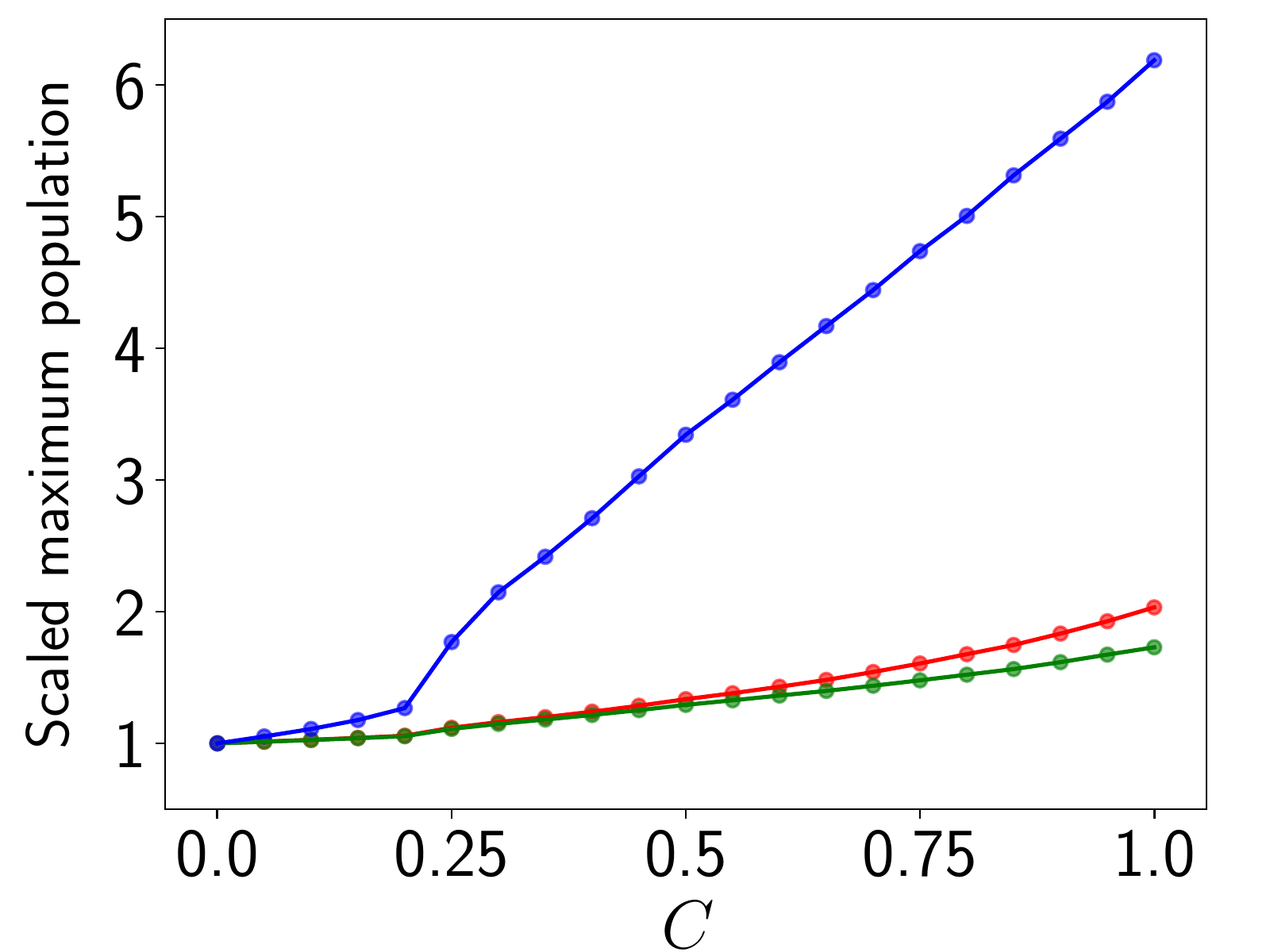}
	\caption{
		Dependence of the global maximum of the vegetation population $u_{max}$ (red), the prey population $v_{max}$ (green) and the predator population $w_{max}$ (blue), occurring in the patches in an interval of time $T=50$, on coupling strength, for $N=100$. Here we depict the {\em scaled} values of the maximum, where $u_{max}$, $v_{max}$ and $w_{max}$ are divided by their values in the uncoupled (i.e. $C=0$) case.
		%(i.e. $u_{max}$ obtained for coupling strength $C$ is divided by the value of $u_{max}$ obtained for $C=0$, and similarly for $v_{max}$ and $w_{max}$). 
		Note that larger systems yield larger $u_{max}$, $v_{max}$ and $w_{max}$, with the values saturating to a characteristic value in the limit of large system size.
%	Dependence of the global maximum $w_{max}$ of the predator population $w_i$ of all the patches $i=1, \dots N$, occurring in an interval of time $T=50$, on coupling strength, for ring sizes $N=10$ (red) and $100$ (blue) in the left panel, and on system size $N$, for coupling strengths $C=0.6$  (red) and $1.0$ (blue) in the right panel.
}
	\label{fig_global_maximum_with_cp}
\end{figure}

Fig.~\ref{fig_global_maximum_with_cp} shows the maximum $u_{max}$, $v_{max}$ and $w_{max}$, for a wide range of coupling strengths of the population patches. In the figure we depict the {\em scaled} values of the maxima, where $u_{max}$, $v_{max}$ and $w_{max}$ are divided by their values in the uncoupled case. These scaled quantities help us assess the increase in the maxima in coupled networks, compared to that obtained in uncoupled patches, allowing us to specifically gauge the coupling-induced effects on the emergent maximum population densities. It is evident from the simulation results that the magnitude of the maximum vegetation and prey populations (i.e. $u_{max}$ and $v_{max}$ denoted by the red and green curves) do not increase much as coupling strength increases. However, the maximum predator population increases very significantly with coupling strength, with $w_{max}$ at $C=1$ exceeding over six-fold the value obtained for uncoupled patches.
%scaling linearly with the coupling strength, after a critical coupling strength of approximately $0.2$. 
%Also note that larger systems yield larger $u_{max}$, $v_{max}$ and $w_{max}$, with the values saturating to a characteristic value in the limit of large system size.
%%(cf. right panel of Fig.~\ref{fig_global_maximum_with_cp}). The saturated value is larger for population patches that are more strongly coupled, and this is consistent with the results in the left panel of Fig.~\ref{fig_global_maximum_with_cp}.

\bigskip
\bigskip

\noindent
{\bf Temporal evolution of Biomass}\\

The next quantity of interest is  the total biomass of the vegetation, prey and predators, denoted by $b_u$, $b_v$ and $b_w$ respectively. This represents a {\em collective dynamical quantity}, and is given at an instant of time $t$ as follows:
\begin{align}
b_u (t) = \sum_{i=1}^N u_{i}(t) \ ,  \  b_v (t) = \sum_{i=1}^N v_{i}(t) \ , \  b_w (t) = \sum_{i=1}^N w_{i}(t)
\label{eq:biomass}
\end{align}
where $N$ is the system size.
Again, we will examine the presence of large excursions from mean-values, as such explosive growth indicate extreme events in time of a collective quantity.

\begin{figure}[htb]
	\centering
	\includegraphics[width=0.485\linewidth,height=0.95\linewidth
	]{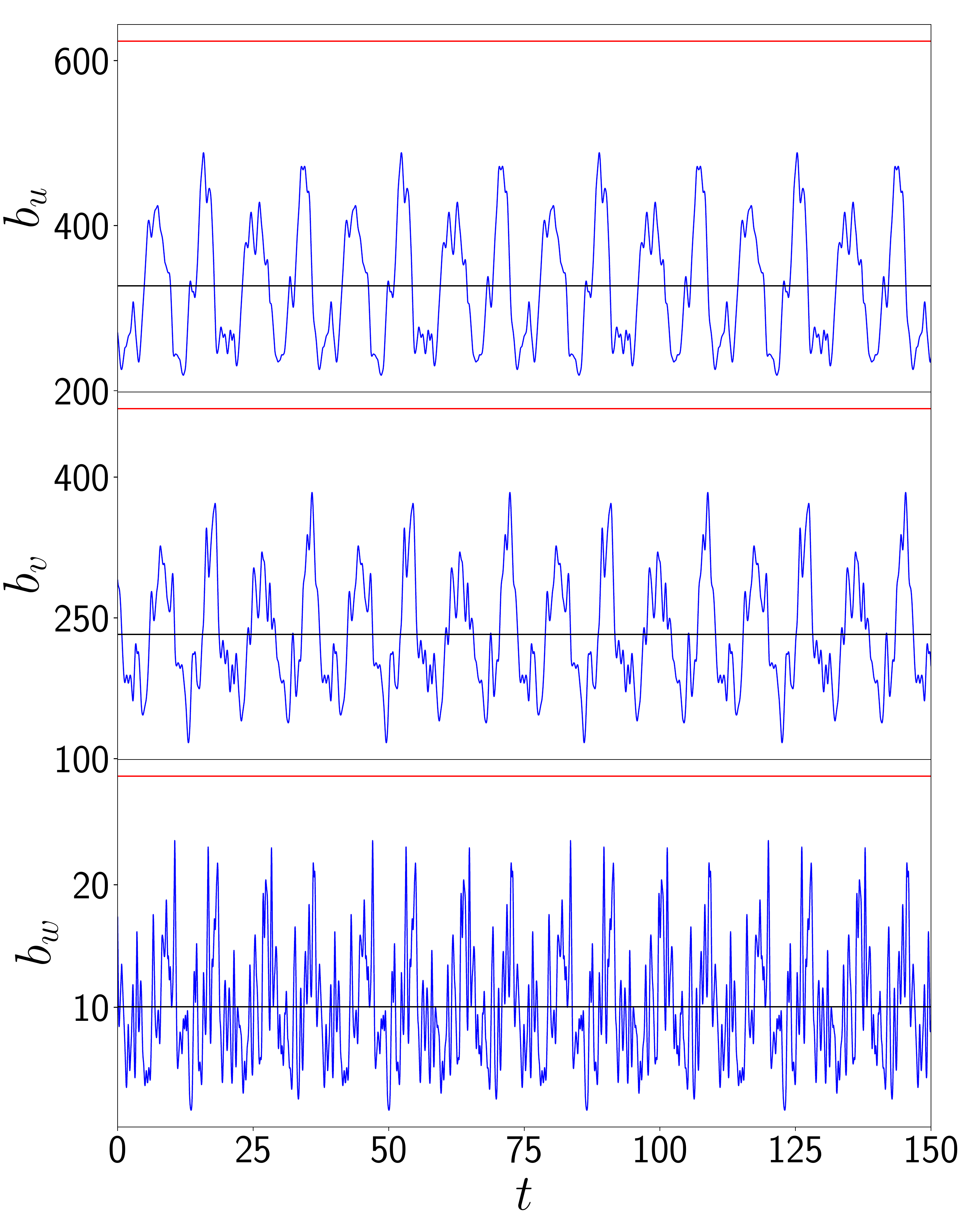}
	\includegraphics[width=0.485\linewidth,height=0.95\linewidth
	]{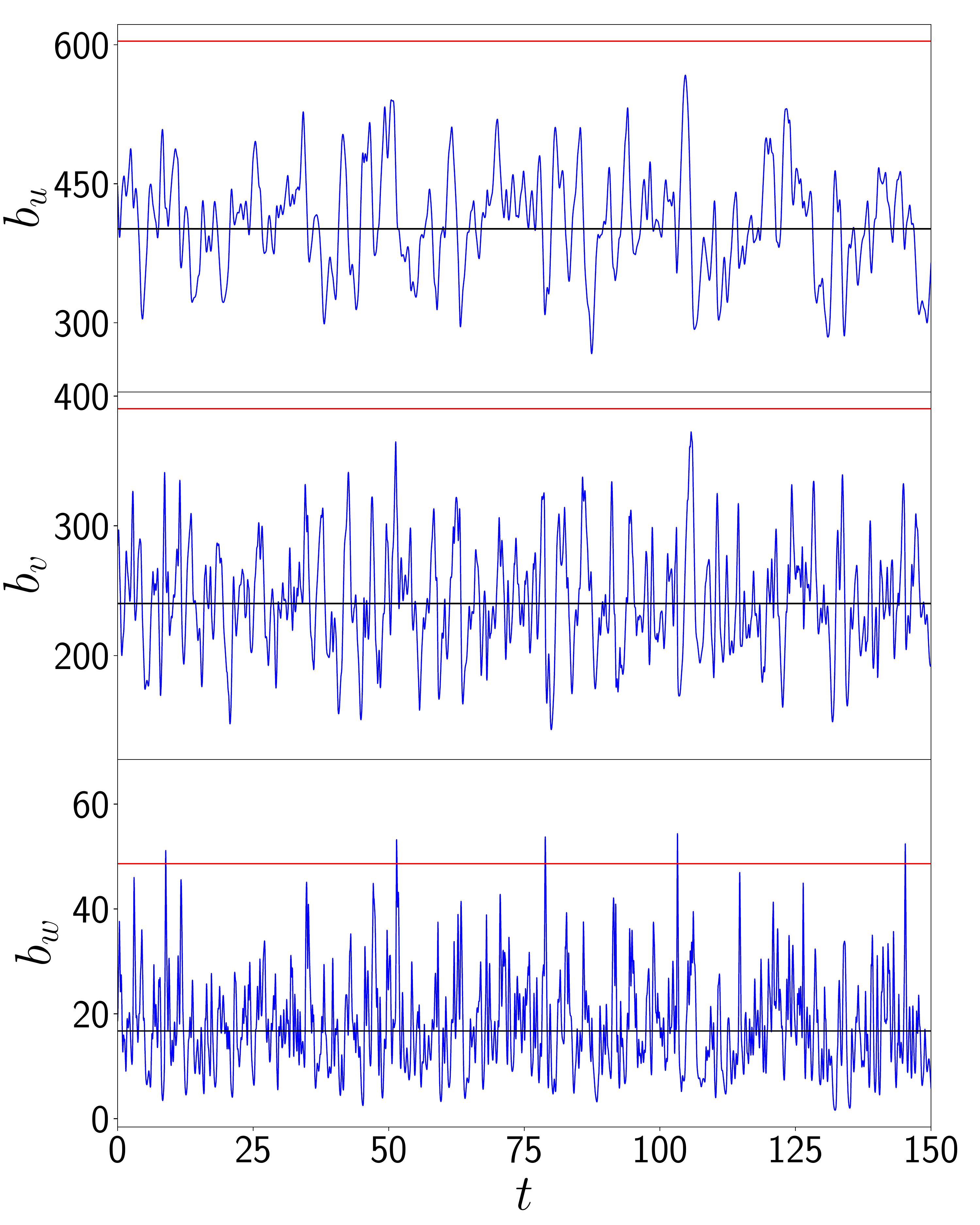}
	\caption{Time evolution of the biomass $b_u (t)$, $b_v (t)$ and $b_w (t)$ of the vegetation, prey and predator populations respectively, for the following cases: (left) uncoupled patches and (right) patches coupled to neighbouring patches, with coupling strength $C=1.0$ in Eqn.~\ref{eqn_model}. The black and red lines represent the mean ($\mu$) and four times the standard deviation $\sigma$ from the mean (i.e. $\mu+4\sigma$) respectively.}
	\label{fig_biomass_with_time}
\end{figure}
%\bigskip

\begin{figure}[htb]
	\centering
	\includegraphics[width=0.55\linewidth]{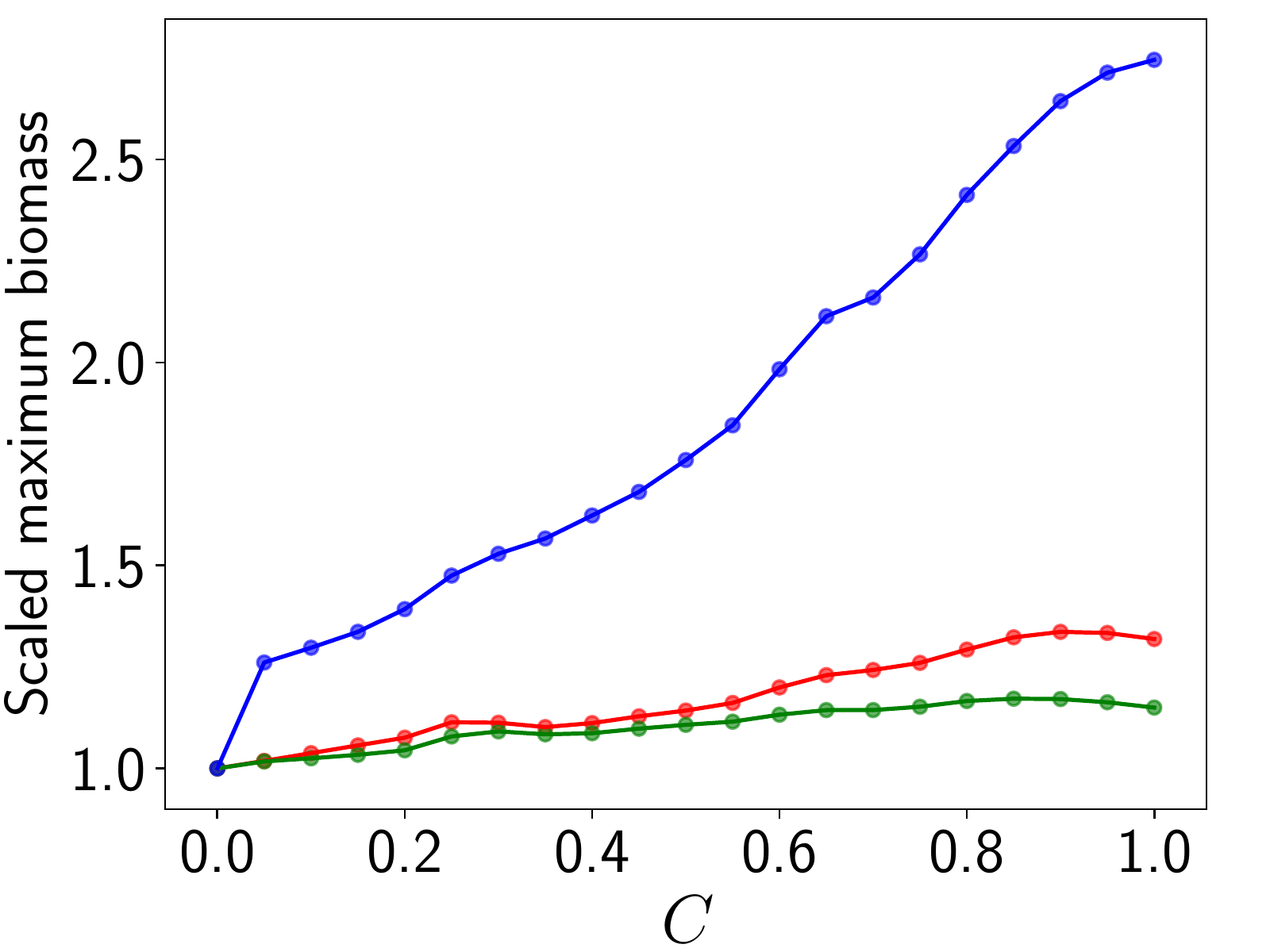}
	\caption{Dependence of the maximum of vegetation biomass $b_u$ (red), the prey biomass $b_v$ (green) and the predator biomass $b_w$ (blue), in an interval of time $T=50$, on coupling strength, for $N=100$. Here we depict the values of the maximum scaled by their values in the uncoupled case.}
		\label{max_biomass}
		\end{figure}

Fig.~\ref{fig_biomass_with_time} shows the biomass of the vegetation ($b_u$), prey ($b_v$) and predators
($b_w$). The uncoupled case is shown alongside as a reference. It is clear that when the patches are uncoupled, $b_u$, $b_v$ and $b_w$ do not experience any large fluctuations. Further, the biomass of vegetation and prey for the coupled case also stays bounded within the $4 \sigma$ threshold. However, interestingly, the predator biomass in coupled patches occasionally builds up to extreme values, crossing the $4 \sigma$ threshold. This is also corroborated through a comparison of the maximum values of the biomass of the vegetation, prey and predator in the coupled network vis-a-vis the uncoupled patches. The maximum prey biomass is almost unchanged on coupling, and the maximum vegetation biomass in a coupled network exhibits
%is approximately $488$ for $C=0$ and $573$ for $C=1$ (i.e. 
less than $20 \%$ change from uncoupled values. 
%The maximum prey biomass also exhibits only around $0.25 \%$ change from uncoupled values.
%is approximately $384$ for $C=0$ and $385$ for $C=1$. 
On the other hand, the maximum predator biomass in a coupled network exhibits a {\em three-fold} increase compared to uncoupled patches (cf. Fig.~\ref{max_biomass}). So coupling has a very significant effect on the predator biomass, and one finds clear evidence of the emergence of extreme events in time for this collective quantity. %reflecting the total biomass of the predators in the entire population network.

\begin{figure}[htb]
	\centering
	\includegraphics[width=0.49\linewidth]{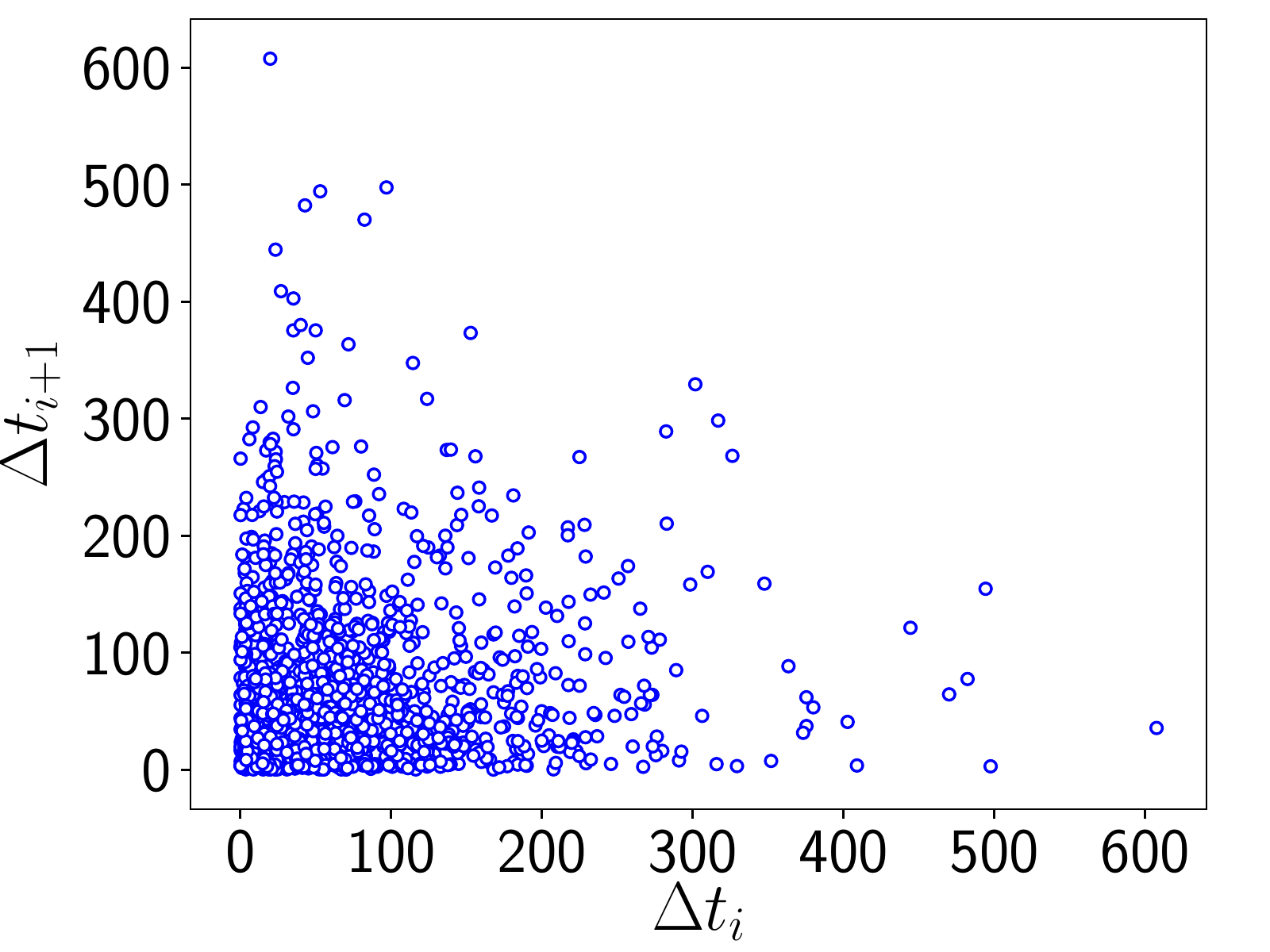} 
%	(a)
	\includegraphics[width=0.49\linewidth]{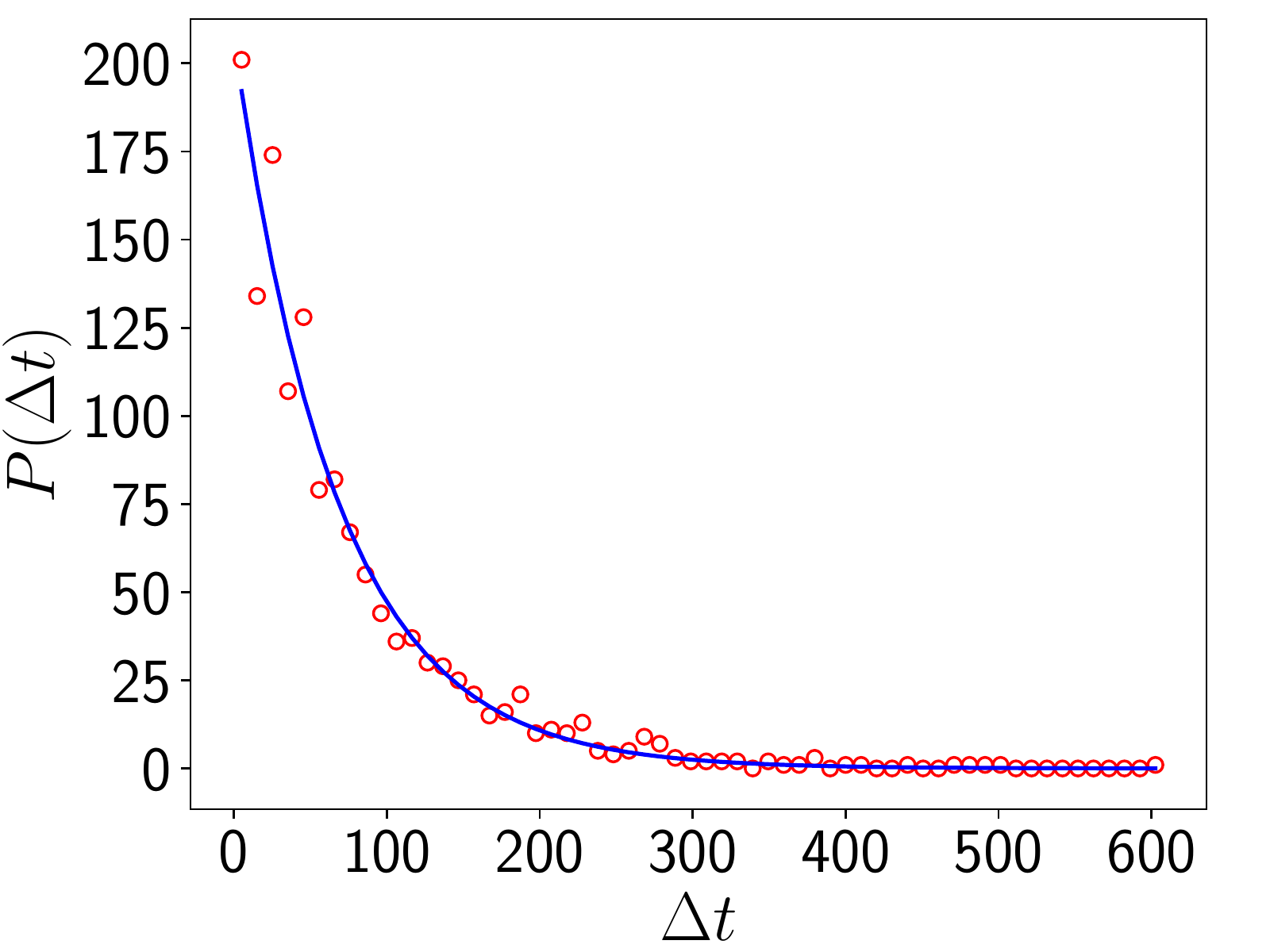} 
%	(b)
	\caption{(Left) Return Map of $\Delta t_{i+1}$ vs $\Delta t_i$, and (right) Probability distribution of $\Delta t_i$ fitted with exponentially decaying function, where $\Delta t_i$ is the $i^{th}$ interval between successive extreme events, where an extreme event is defined at the instant when biomass crosses the $\mu+4\sigma$ line (cf. Fig~\ref{fig_biomass_with_time}). Here the system size $N=100$ and coupling strength $C=1.0$ in Eqn.~\ref{eqn_model}.}
	\label{fig_del_t_of_peak_crossing_4sigma}
\end{figure}

In order to ascertain that the extreme values are uncorrelated and aperiodic we examine the time intervals between successive extreme events in the predator biomass evolution. %Note that for biomass we consider the $4 \sigma$ threshold for extreme events.  
Fig.~\ref{fig_del_t_of_peak_crossing_4sigma} shows the return map of the intervals between extreme events and it is clearly shows no regularity. The probability distribution of the intervals is also Poisson distributed and so the extreme predator population buildups are uncorrelated aperiodic events.

%%% For this one (w_max vs. N), also put \varepsilon=0 (no coupling case) for comparison
%\begin{figure}[H]
%	\centering
%	\includegraphics[width=1\linewidth]{wmax_vs_RingSize_cp_0.6_1.0.pdf}
%	\caption{Global maximum of predator population of all patches with varying ring size ($N$), for coupling strengths $\varepsilon=0.6$ (red) and $1.0$ (blue).}
%	\label{fig_global_maximum_with_N}
%\end{figure}

\bigskip
\bigskip

\noindent
\textbf{Generalized Extreme Value distribution}\\

Lastly we analyse the distribution of the maximum size of the predator populations, as well as the maximum size of the collective predator biomass. In order to obtain a quantitative measure of the extreme events generated for different coupling strengths, we fit these probability distributions to the probability density function of the generalized extreme value distribution, given by:

\begin{eqnarray*}
    f(y;\zeta)= 
\begin{cases}
	\exp(-(1-\zeta y)^{1/\zeta })(1-\zeta y)^{(1/\zeta) -1},& \zeta >0,\\ \ & \ y\leq 1/\zeta\\
	\exp(-\exp(-y))\exp(-y),              & \zeta =0
\end{cases}
\label{gev}
\end{eqnarray*}

with 
\begin{eqnarray*}
y(x,\mu,\sigma)&=&(x-\mu)/\sigma
\end{eqnarray*}

Here $\zeta$ is the shape parameter, $\mu$ is the location parameter and $\sigma$ is the scale parameter. Note that the scale parameter is the most relevant parameter in this case as it determines the spread of the distribution. 

\begin{figure}[htb]
	\centering
	\includegraphics[width=0.8\linewidth,height=0.4\linewidth]{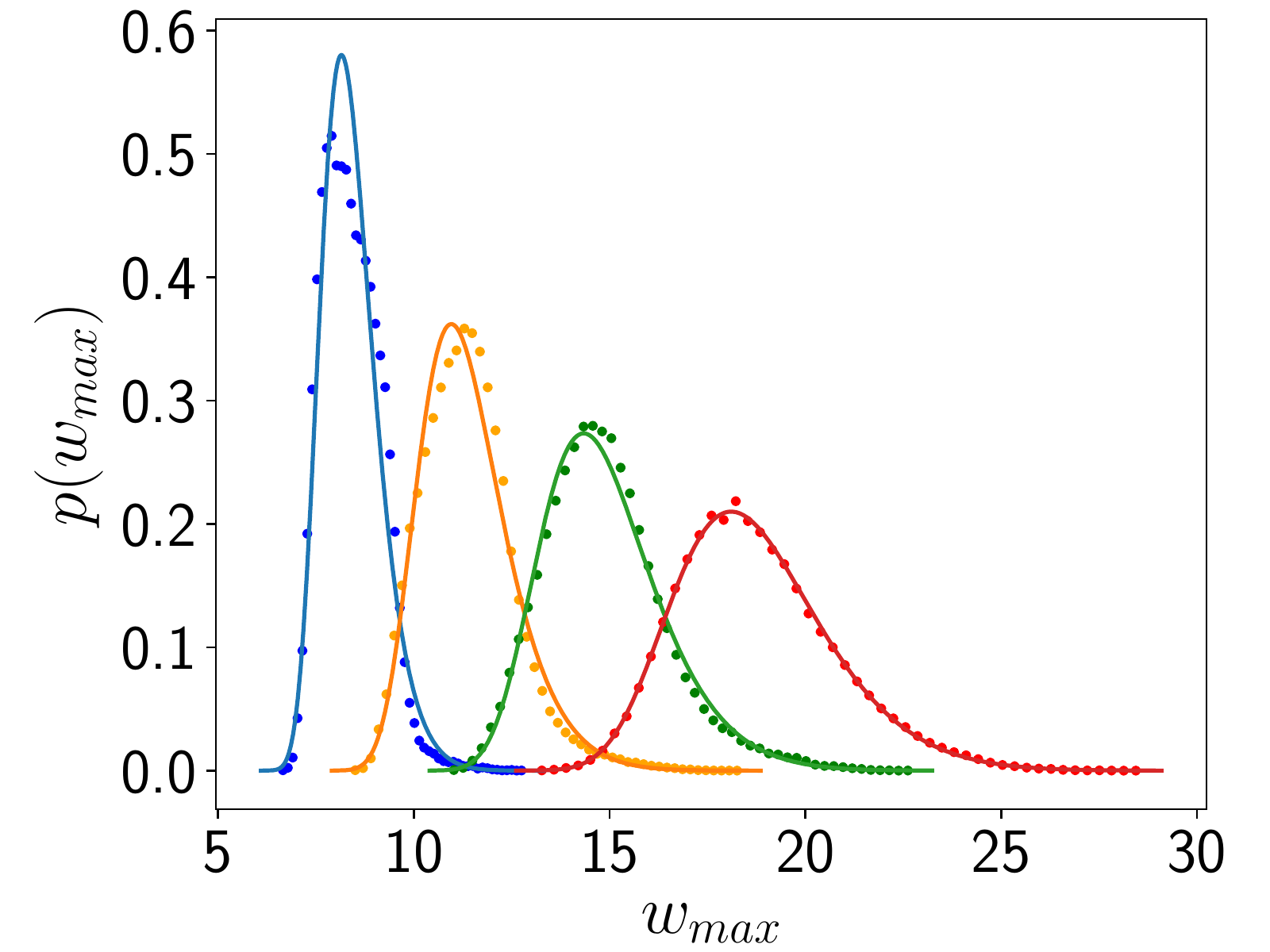}
\includegraphics[width=0.8\linewidth,height=0.4\linewidth]{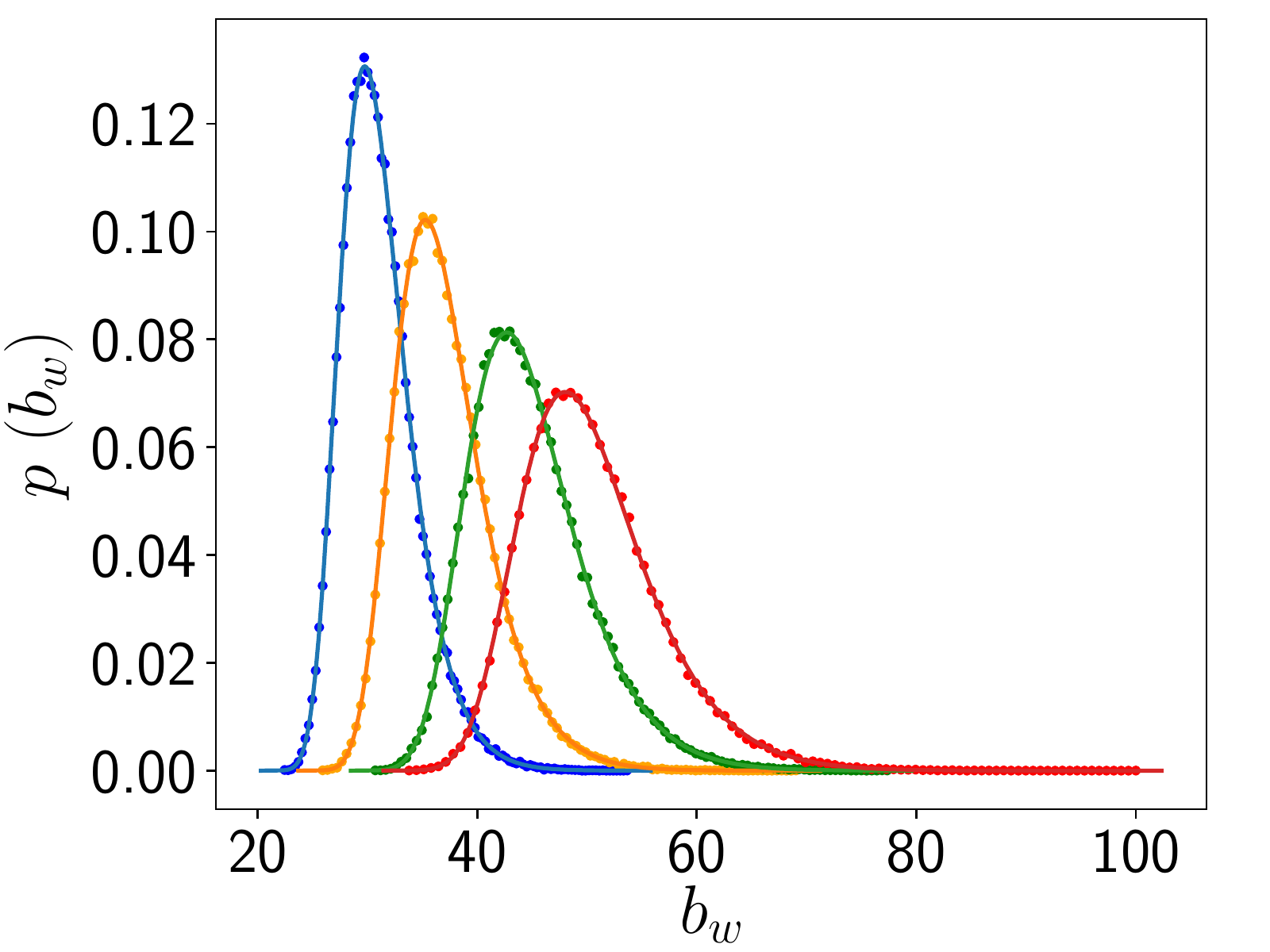}
		\caption{Probability distribution of the (top) maximum predator population in a patch $w_{max}$, and (bottom) maximum predator biomass, obtained from sampling the predator populations in a time interval $T=50$ (with no loss of generality). The fit of the data from numerical simulations to the Generalized Extreme Value distribution (cf. Eqn.~\ref{gev}) is shown by solid lines, 
%for coupling strength $C=0.2$ (blue), $0.4$ (orange), $0.6$ (green), $0.8$ (red) and $1.0$ (purple) in panel (a),
%	\label{fig_gev_fitting_of_biomass}
%Probability distribution of the maximum predator population in a patch $w_{max}$, obtained from sampling the predator populations in a time interval $T=50$. The fit of the data from numerical simulations to the Generalized Extreme Value distribution (cf. Eqn.~\ref{gev}) is shown by solid lines, 
for coupling strength $C=0.4$ (blue), $0.6$ (orange), $0.8$ (green) and $1.0$ (red).
}
	\label{fig_gev_fitting_of_max_predator_population}
\end{figure}

\begin{figure}[htb]
	\centering
\includegraphics[width=0.35\linewidth]{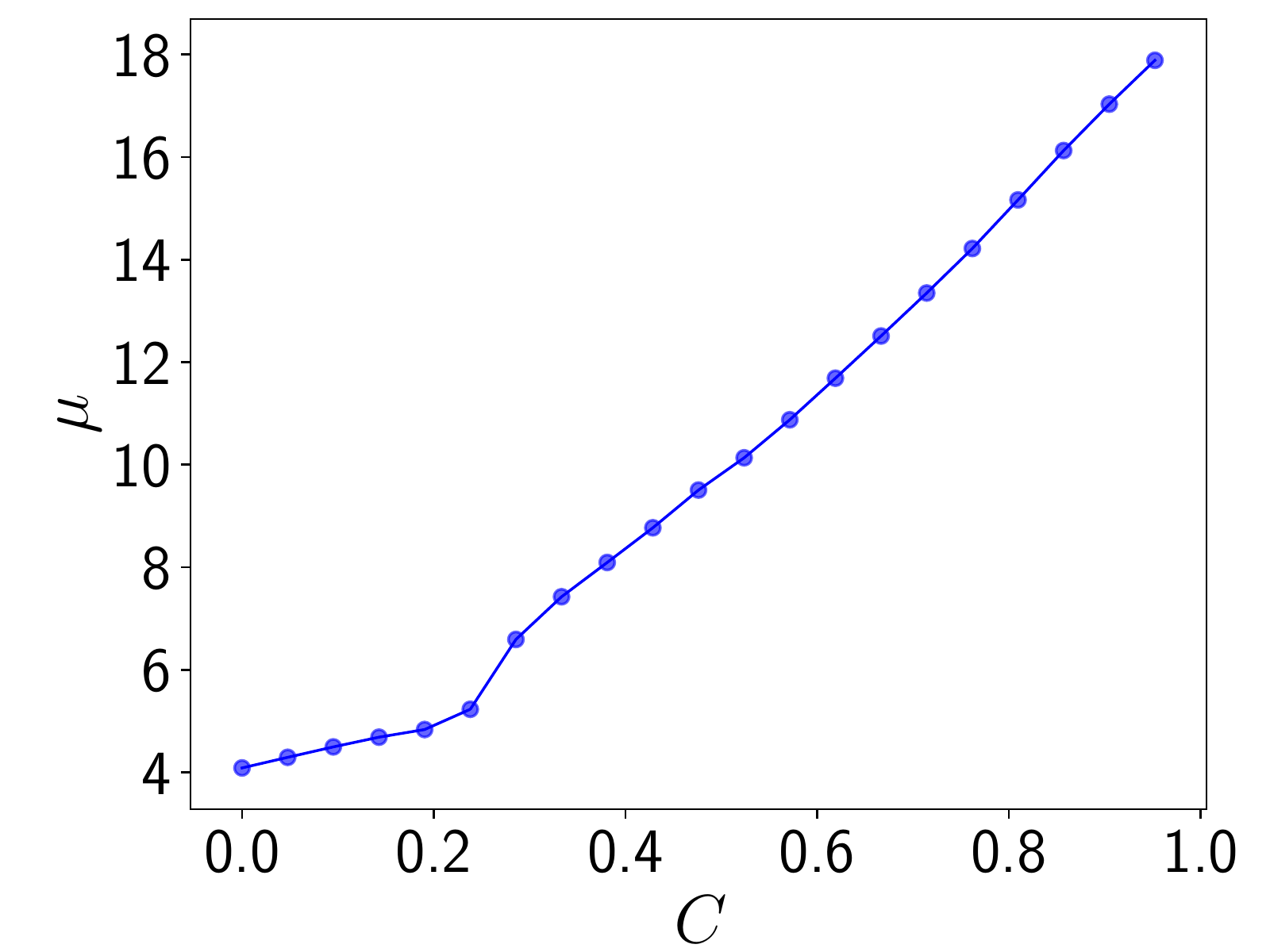}
	\includegraphics[width=0.35\linewidth]{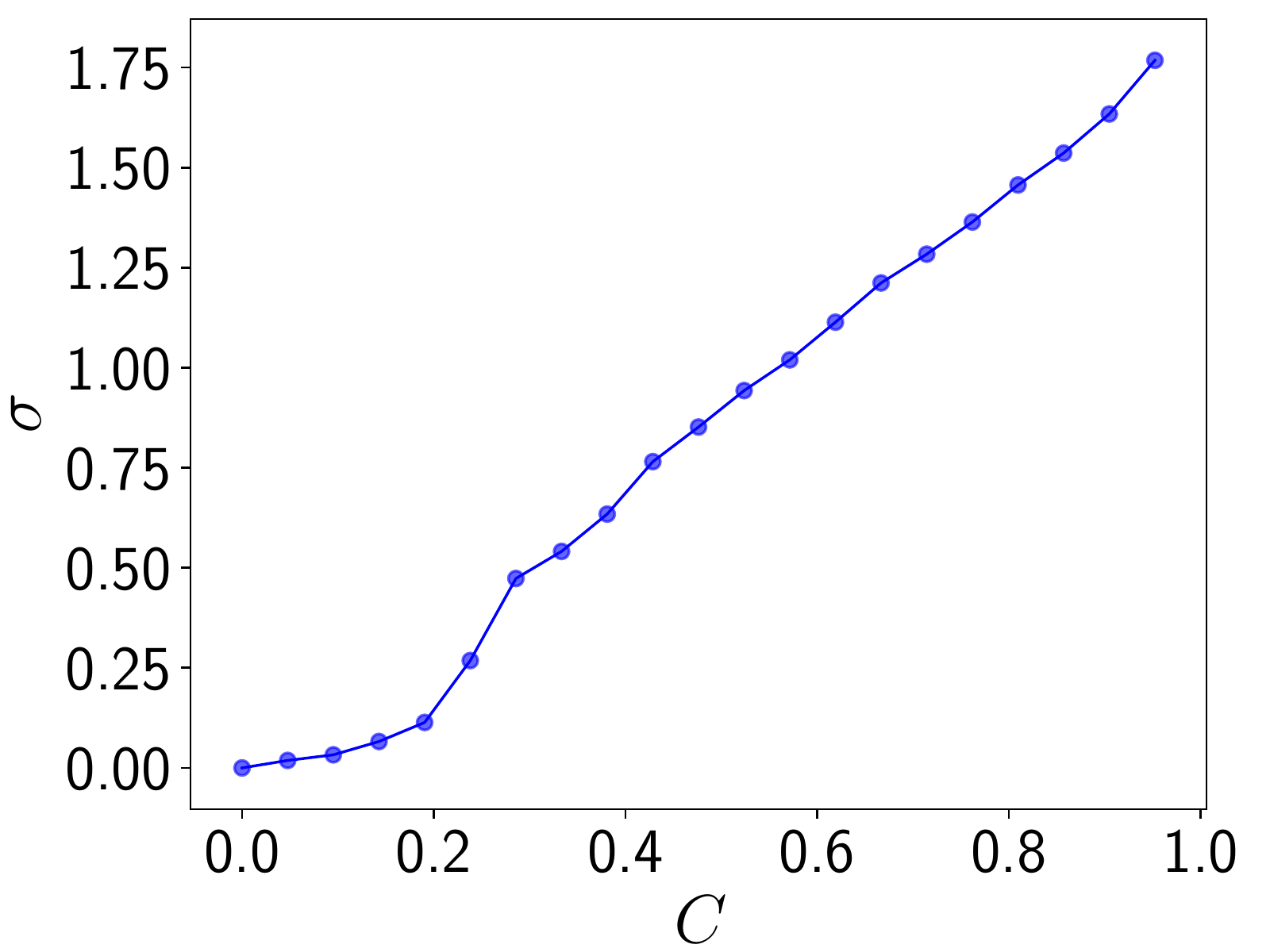}
%	
%	(a) 
%	
	\includegraphics[width=0.35\linewidth]{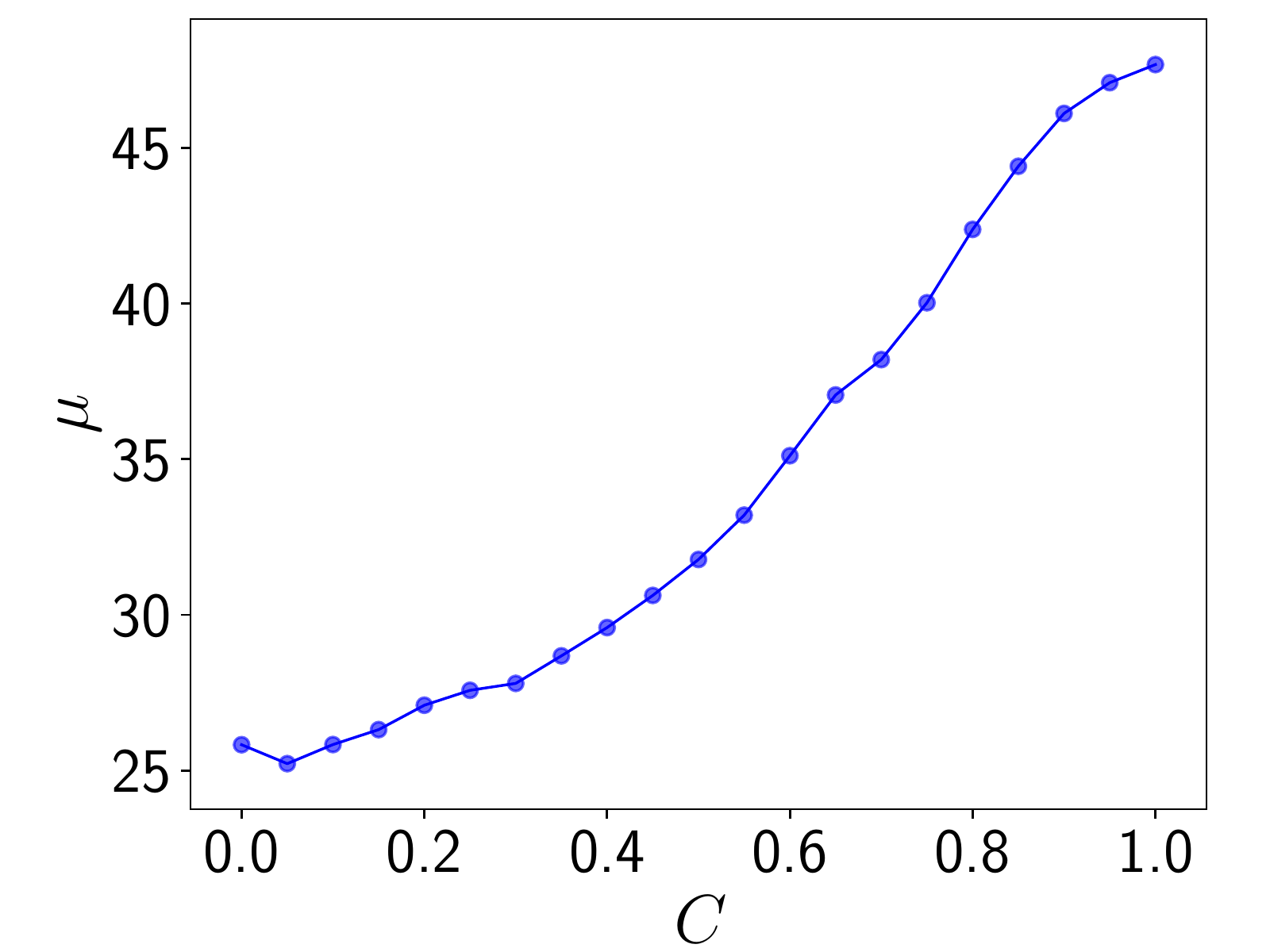}
	\includegraphics[width=0.35\linewidth]{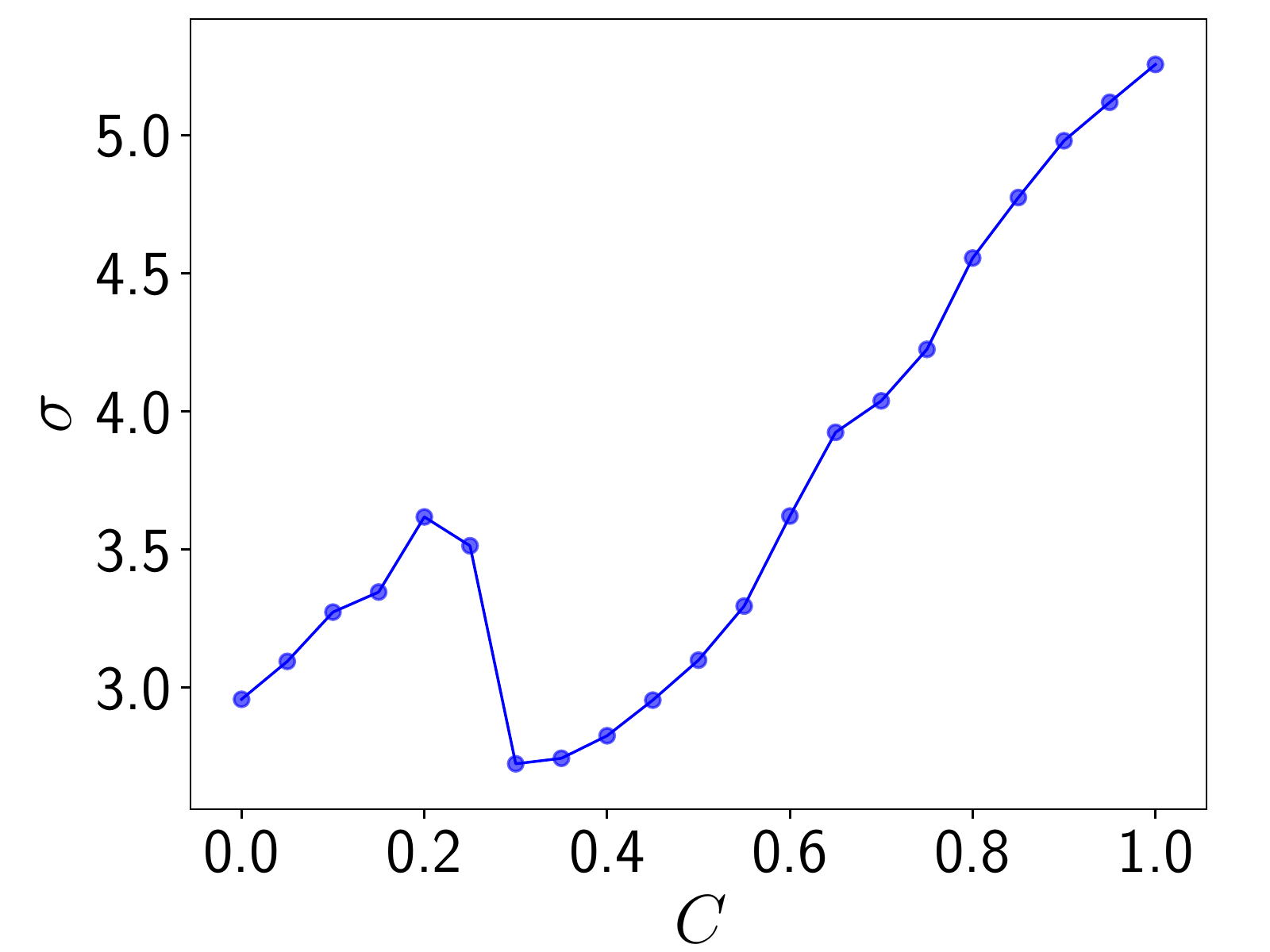}
%
%	(b)
\caption{Location (left) and scale (right) parameters obtained by best-fit to the Generalized Extreme Value distribution of (top) maximum predator population in a patch $w_{max}$, and (bottom) maximum predator biomass $b_w$, obtained from sampling a time interval of $T=50$ (cf. Fig~\ref{fig_gev_fitting_of_max_predator_population}), for different coupling strengths $C$.}
	\label{fig_mu_and_sigma_from_gev_fitting_of_max_predator_population}
\end{figure}

Fig.~\ref{fig_mu_and_sigma_from_gev_fitting_of_max_predator_population} shows the location and scale parameters obtained by best-fit to the Generalized Extreme Value distribution of the maximum predator population in a patch $w_{max}$, and the maximum predator biomass $b_w$. Clearly the location and scale parameters increase monotonically with coupling strength, for coupling strengths higher than $\sim 0.4$, with the rise being approximately linear at high $C$. Increasing location parameters indicate that the average predator population in a patch and the average predator biomass increase almost linearly with coupling strength. Increasing scale parameters suggest that the distribution becomes increasingly spread out, and the tail of the distribution extends to larger values. So more extreme predator populations can be expected to occur in the patches, from time to time, when the coupling between the patches is stronger.

\bigskip
\bigskip

\noindent
{\bf Conclusions}\\

In summary, we have studied the population dynamics of a ring of patches with vegetation, preys and predators. The population dynamics in the patches is given by a model for the snowshoe hare and the Canadian lynx that fits observed data well, and the patches are coupled through interactions of the Lotka-Volterra type. We find that this system yields extreme events in the predator population in the patches, with bursts of explosive  predator population growth in a few isolated patches from time to time. Further, the collective predator biomass also yields extreme values as the coupled system evolves. The maximum value of the predator population in a patch, as well as the maximum value of the predator biomass, increases with coupling strength, indicating the importance of coupling in the generation of such extreme events. Fits of the data from numerical simulations to Generalized Extreme Value distributions also quantitatively corroborate these trends.

Our results then are important, both from the view-point of general models of coupled nonlinear systems, as well as for the more specific implications it may potentially hold for population dynamics. So first we have demonstrated how a {\em deterministic system, with generic Lotka-Volterra type of interactions, can give rise to extreme events in space and time}. Such examples are uncommon, and so they are  significant in the context of general complex systems. Secondly, in the specific context of population dynamics, our model system suggests how predator population densities can grow explosively in certain patches. Though relatively rare, the magnitude of these extreme bursts of predator population density is so huge, that the damage or ensuing cost to control the event, is considerable. Further the biomass of predators can also grow extremely large at certain points in time, and this is of significance due to the  catastrophic effects large predator populations can have on the ecosystem as a whole. Lastly, interestingly, the Lotka–Volterra class of interactions has also been used extensively in economic theory \cite{economic}, and so our results may have some bearing on extreme events in the financial context.\\

%	\bibliographystyle{unsrt}
%	
%	\bibliographystyle{amsalpha}
%	\bibliography{my_bib}

\end{document}